\documentclass[a4paper]{article}
\usepackage{graphicx}
\usepackage{amsmath}

\begin{document}

\begin{center}
{\bf \Large
Pores in a two-dimensional network of DNA strands - computer simulations
}\\[5mm]

{\large
M. J. Krawczyk and K. Ku{\l}akowski
}\\[3mm]

{\em
Department of Applied Computer Science,
Faculty of Physics and Nuclear Techniques,
AGH University of Science and Technology\\
al. Mickiewicza 30, PL-30059 Krak\'ow, Poland
}

\bigskip
{\tt krawczyk@novell.ftj.agh.edu.pl, kulakowski@novell.ftj.agh.edu.pl}

\bigskip
\today
\end{center}

\begin{abstract}
Formation of random network of DNA strands is simulated on a two-dimensional triangular 
lattice. We investigate the size distribution of pores in the network.
The results are interpreted within theory of percolation on Bethe lattice.
\end{abstract}



\section{Introduction.}
Besides their extremely important biological function, molecules of DNA are useful in many areas, 
as biotechnology, nanotechnology and electronics.  Possibility of design of molecules 
of exactly known composition and length opened an area for using DNA for construction 
of different structures. Appropriate choice of DNA sequences allows to obtain not 
only linear chains but also branched DNA molecules. These synthetical branches 
are analogs of branched DNA molecules (Holliday junctions) discovered in living cells.
Addition of sticky ends to such junctions enables a construction of complex molecular systems. 
Holliday junctions occurring naturally are present between two symmetrical segments 
of DNA strands, and because of that, the branch point is free to migrate. Such a behaviour is not 
advantageous when DNA strands are used for a construction of networks. Elimination of symmetry 
in synthetically made DNA molecules fixes the branch points location. Synthetical junctions give 
an opportunity for building complex structures from branches prepared specially for given purposes; 
this enables to construct junctions with different numbers of arms. Addition of sticky ends 
to three- or four-armed branches allows to construct two-dimensional lattices, and using 
six-armed branches -- for a construction of 3D lattices \cite{see,see1}. Different kinds of 
branched molecules can be connected with linear strands or to each other. Appropriate choice of DNA and other 
allows for a construction of lattices of specific parameters.\\
One of the most important problems occurring during the network formation is that the angles 
between the arms of branched junctions are variable.  This problem can be solved by using 
branches with four different sticky ends. Besides a stabilization of the edges, this solution 
eliminates improper connections between arms. 
Second important problem with DNA network formation is that the polymer strands are flexible.
They can cyclize on itself what prevents further growth of the system \cite{see1}. 
To construct periodic lattices, such as e.g. crystals, which can be used in 
nanoelectronics, polymers must be rigid. Construction of crystals is possible since 
 DNA double-crossover motifs were developed \cite{see1}.  
As the DNA double helix width is about 2 nm and one helical repeat length is about 3.4 nm, 
lattices made from DNA strands can be useful in the emerging nanotechnology. It is possible 
to use DNA to form crystals with edge length of more than 10 $\mu$m and a uniform thickness of 
about 1-2 nm \cite{nie}. One of main advantages of DNA structures is a possibility of their 
modifications by binding different ions and groups of ions or fluorescent labels \cite{ito,see1}. 
This is useful in electronic devices where nanometer-scale precision is extremely important.\\

Here we are interested in a design of random networks of DNA molecules. This aim is motivated as 
follows: the above mentioned limitations and conditions of the production of DNA lattices and 
crystals can be found to be too severe and in consequence too costful for some applications,
where periodic structures are not necessary. In this case, properties of networks of DNA can be
relevant for these potential applications of various two-dimensional structures, e.g. 
of polymers and liquid crystals \cite{a1,a2}. On the other hand, process of formation of a 
random network is close to some model processes of current interest for theories of disordered
matter \cite{t1}. Then, interpretation of our numerical results allows to comment to what extent 
assumptions of these theories are minimal. As it is discussed below, our results suggest that some 
conclusions of the percolation theory in uncorrelated Bethe lattice can apply to correlated 
structures.

We are going to investigate the size distribution of empty holes (pores) in a random network. This aim
should allow to evaluate the network in a role of a kind of molecular sieve \cite{c}. Whilst 
disorder in such a sieve is obviously a disadvantage when compared to periodic structures 
\cite{see1}, it seems to us worthwhile to evaluate consequences of this disorder.

In subsequent section the model is described and some details are given on its numerical 
realization. In Section III we show the results on the size distribution of the pores, obtained 
for a triangular two-dimensional lattice. Short discussion closes the text.

\section{Model}
In our model we initially consider two kinds of molecules: linear DNA strands and DNA-junctions, 
which connect ends of the strands. The latter can be of three or four arms; they are
called Y- and X-junctions. We assume that the length of the linear molecules is much larger, than the size of the 
junctions. 
Also, we assume that the number of junctions is so much larger than the number of strands,
that \emph{i)} practically any collision of two ends of the strands leads to their fixed 
connection,
\emph{ii)} the junctions do not influence the motion of the strands in other way, than as in 
\emph{i)}. Summarizing, our assumptions on the DNA-junctions (enough small and enough numerous) 
allow to neglect them during the simulation.

To speed up numerical calculations, the algorithm is based on a periodic 
lattice. It is straightforward to apply the square lattice in the case of X-junctions, and
the triangular lattice in the case of Y-junctions. 
The length of the linear DNA strand is chosen to be an integer multiple of the size of the lattice 
cell. An additional assumption is that, no more than arbitrary chosen number 
$M$ of different DNA strands are placed at the same lattice node, at the same time.
This limits the density of the system from above, to reproduce the steric effect.
We have only three 
model parameters: the length $L$ of the strands, their number $N$, and the maximal number
$M$ of strands at one lattice cell.
A strand cannot rotate, 
but only moves along its axis. The direction of the latter is random, with equal 
probabilities 1/2 at each time step.  
Initial set of the strand positions is selected to be entirely random (positions and 
orientations), with only the above 
mentioned limitation of the density. We choose periodic boundary conditions, to limit an 
influence of the borders. Subsequent positions are controlled 
by the assumption that in the one cell can not be more molecules that arbitrary chosen 
number M.  Possible orientations of the strands are horizontal or 
vertical in the case of square lattice. In the case of triangular lattice, 
three orientations of the strands are equally possible, along all three sides of the triangle. 
If the 
number of ends of the strands at a given cell of the lattice is equal to or more than two (but 
not more than $M$, which is forbidden by the code), these strands become connected to a 
DNA-junction and to each other and they cannot move anymore. If at any time step two different 
strands are placed at the same cells only one of them, chosen randomly, 
can be connected to the DNA-junction. 
In one 
simulation step all molecules which are not immobile can move with equal probability to 
the right or left if they are oriented horizontally, and up or down if they are vertical 
in the case of square lattice. In the case  of triangular lattice the strands can move in
two possible directions, in accordance with their orientation. At each time step each strand 
can move by the length equal to the size of the one lattice cell. A movement is accepted 
if, as its consequence, the number of strands at all lattice cells is not more then $M$. 

All results are obtained as average value from ten independent simulations, and the lattice 
size was 512$\times$512. Each simulation was made for 100 time steps. Duration of the simulation 
was chosen so as to a number of free strands was in range of few percent (2-3).
In all simulations we studied distribution of the pores, i.e. the lattice nodes which are not 
occupied by the DNA strands as a function of the model parameters. The size distribution of the 
pores was investigated using the Hoshen-Kopelman algorithm \cite{hos}.

\section{Results} 
The results presented in Figs. 1-4 are obtained for the case of triangular lattice with $M=3$.  
The number $N$ of molecules introduced to the system varies from $10^4$ (Fig. 1) to $4\times10^4$ 
(Fig. 4). The plot shown in Fig. 1 shows that there are large pores in the system, but their  
number is exactly one for each size $s$; in other words, the system is diluted. For larger  
$N$ we see that the pore size distribution is close to a power law, i.e.  
$N_g(s)\propto s^{-\tau}$, with $\tau \approx 1.6$ for $N=2\times10^4$ and $\tau \approx 2.0$ 
for $N=3\times10^4$. This is seen in Figs. 2 and 3.  
A further increase of the number $N$ of molecules leads to a cutoff of the scale-free character  
of the plot (Fig. 4). The size of maximal pores is then reduced from 100 or more to about 30.  

Similar results are obtained for other strand lengths $L$. As a rule, for low values of $N$  
the shape of the function $N_g(s)$ is close to the one in Fig. 1. For larger $N$ we see a power-law 
function, as in Figs. 2 and 3. When $N$ increases further, large pores vanishes. This common  
behaviour is summarized in Fig. 5, where a kind of phase diagram is presented. Assuming that  
the scale-free power-law function is proper at only one value of $L_c(N)$, where the concentration 
of the strands is at the percolation threshold, we marked approximate position of this threshold 
with a schematic solid line. This plot is qualitative: the only conclusion from our data 
is that $L_c$ decreases with $N$. 

\begin{figure}
\vspace{0.3cm}
{\par\centering \resizebox*{9cm}{7cm}{\rotatebox{270}{\includegraphics{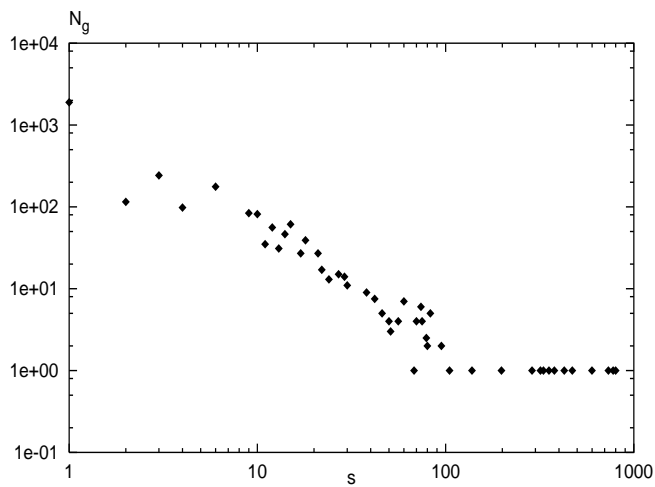}}} \par}
\vspace{0.3cm}
\caption{Pore size distribution $N_g(s)$ for $L=7$, $N=10^4$, $M=3$.}
\end{figure}

\begin{figure}
\vspace{0.3cm}
{\par\centering \resizebox*{9cm}{7cm}{\rotatebox{270}{\includegraphics{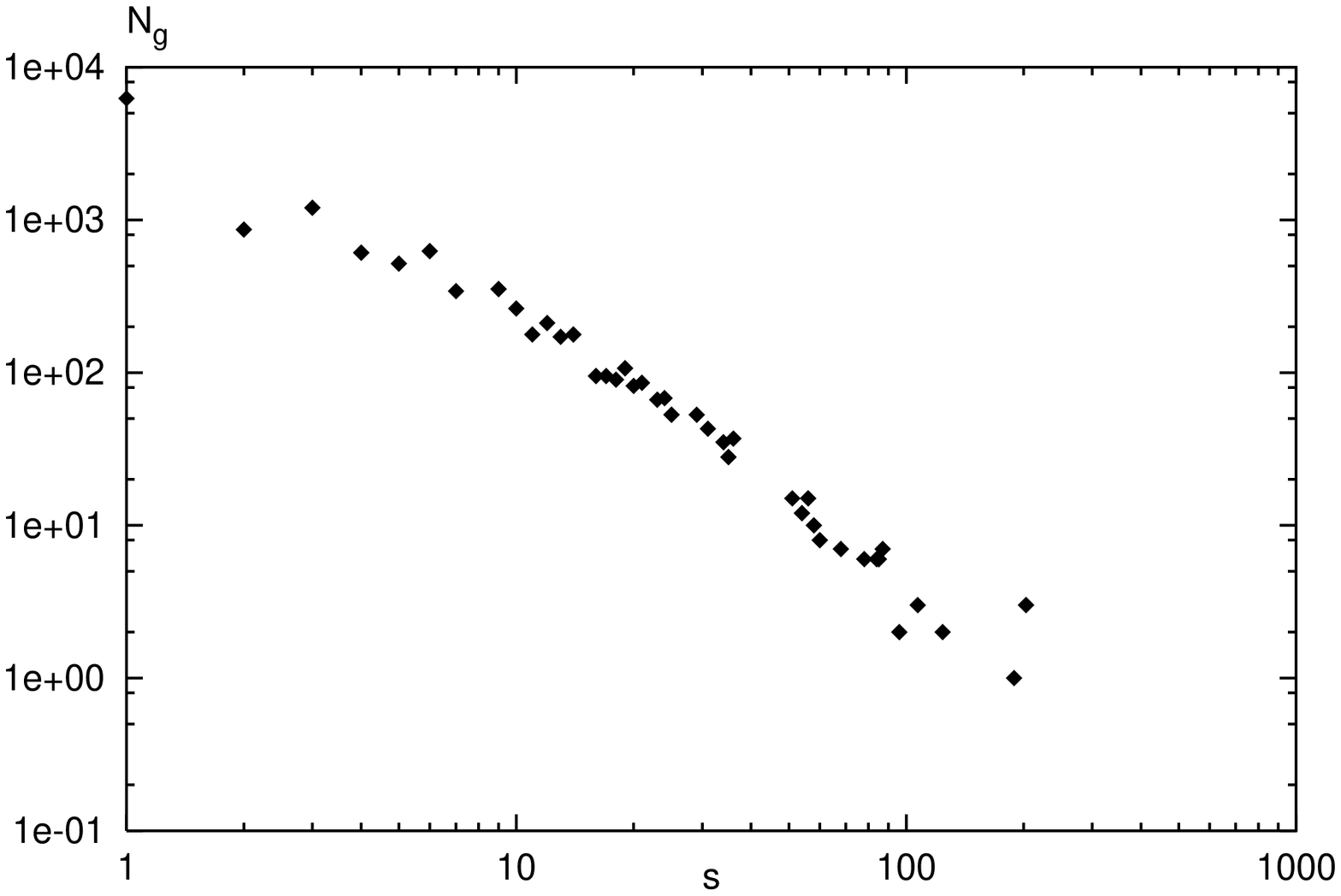}}} \par}
\vspace{0.3cm}
\caption{Pore size distribution $N_g(s)$ for $L=7$, $N=2\times10^4$, $M=3$.}
\end{figure}

\begin{figure}
\vspace{0.3cm}
{\par\centering \resizebox*{9cm}{7cm}{\rotatebox{270}{\includegraphics{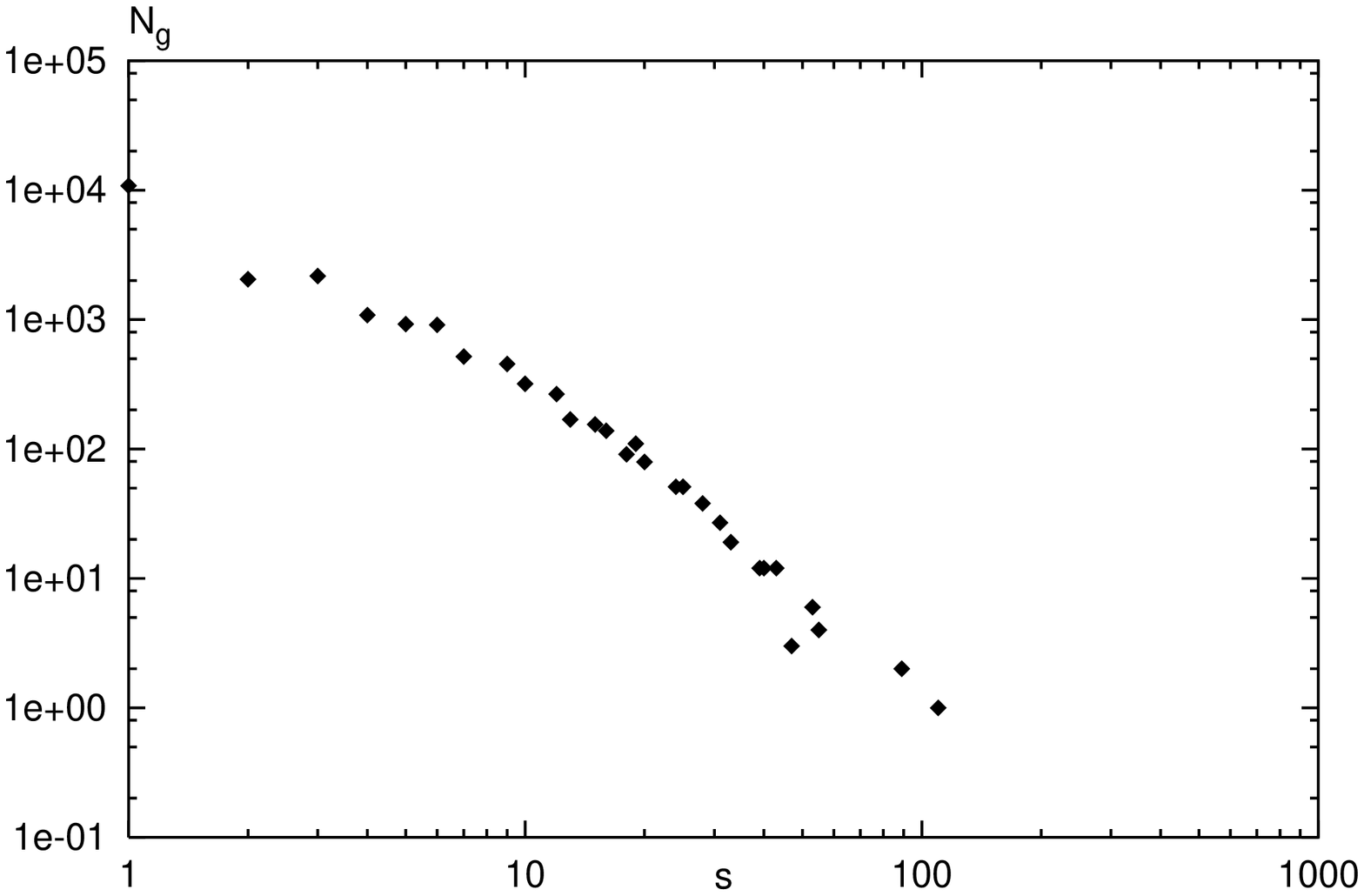}}} \par}
\vspace{0.3cm}
\caption{Pore size distribution $N_g(s)$ for $L=7$, $N=3\times10^4$, $M=3$.}
\end{figure}

\begin{figure}
\vspace{0.3cm}
{\par\centering \resizebox*{9cm}{7cm}{\rotatebox{270}{\includegraphics{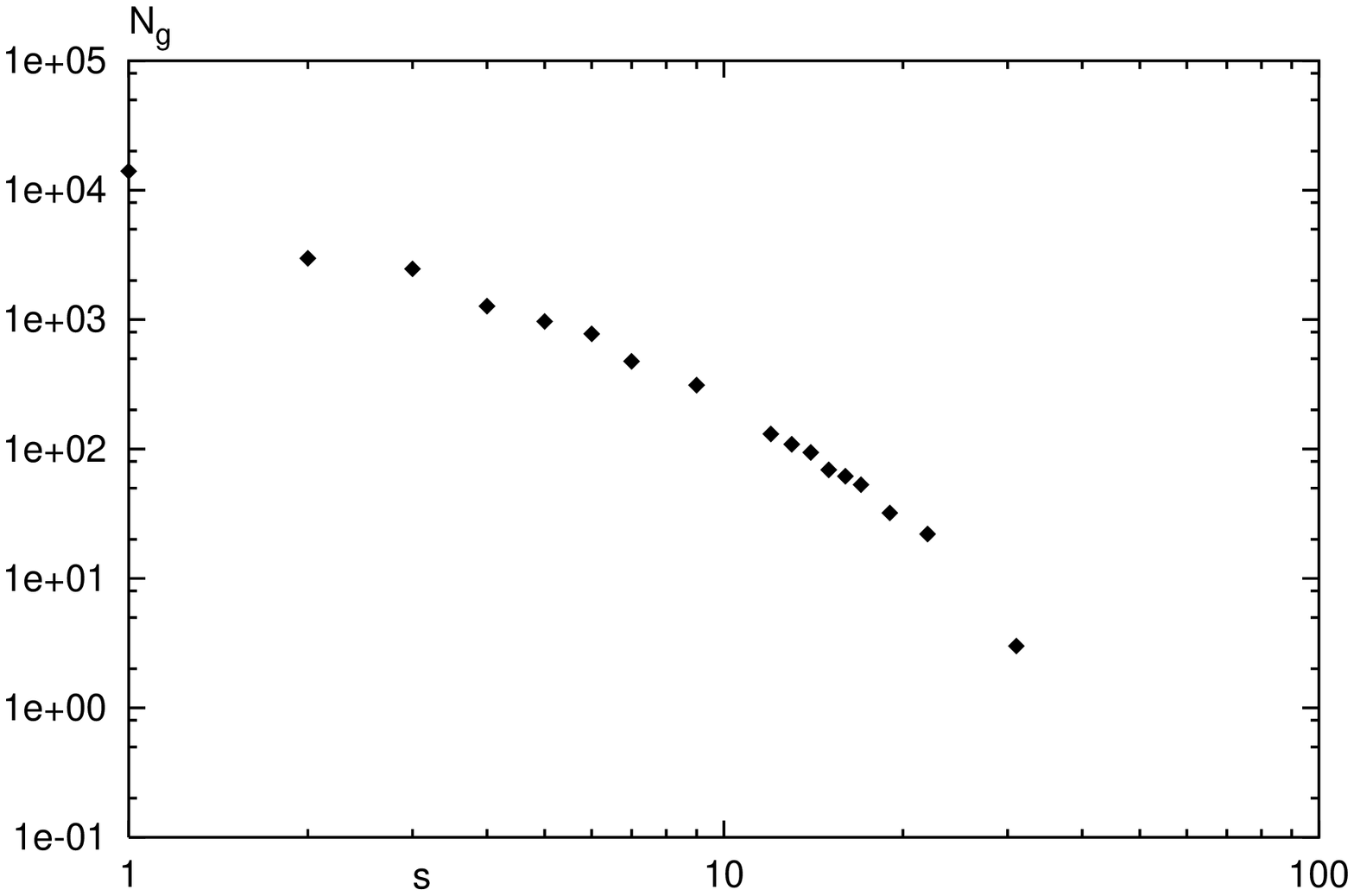}}} \par}
\vspace{0.3cm}
\caption{Pore size distribution $N_g(s)$ for $L=7$, $N=4\times10^4$, $M=3$.}
\end{figure}

\begin{figure}
\vspace{0.3cm}
{\par\centering \resizebox*{9cm}{7cm}{\rotatebox{270}{\includegraphics{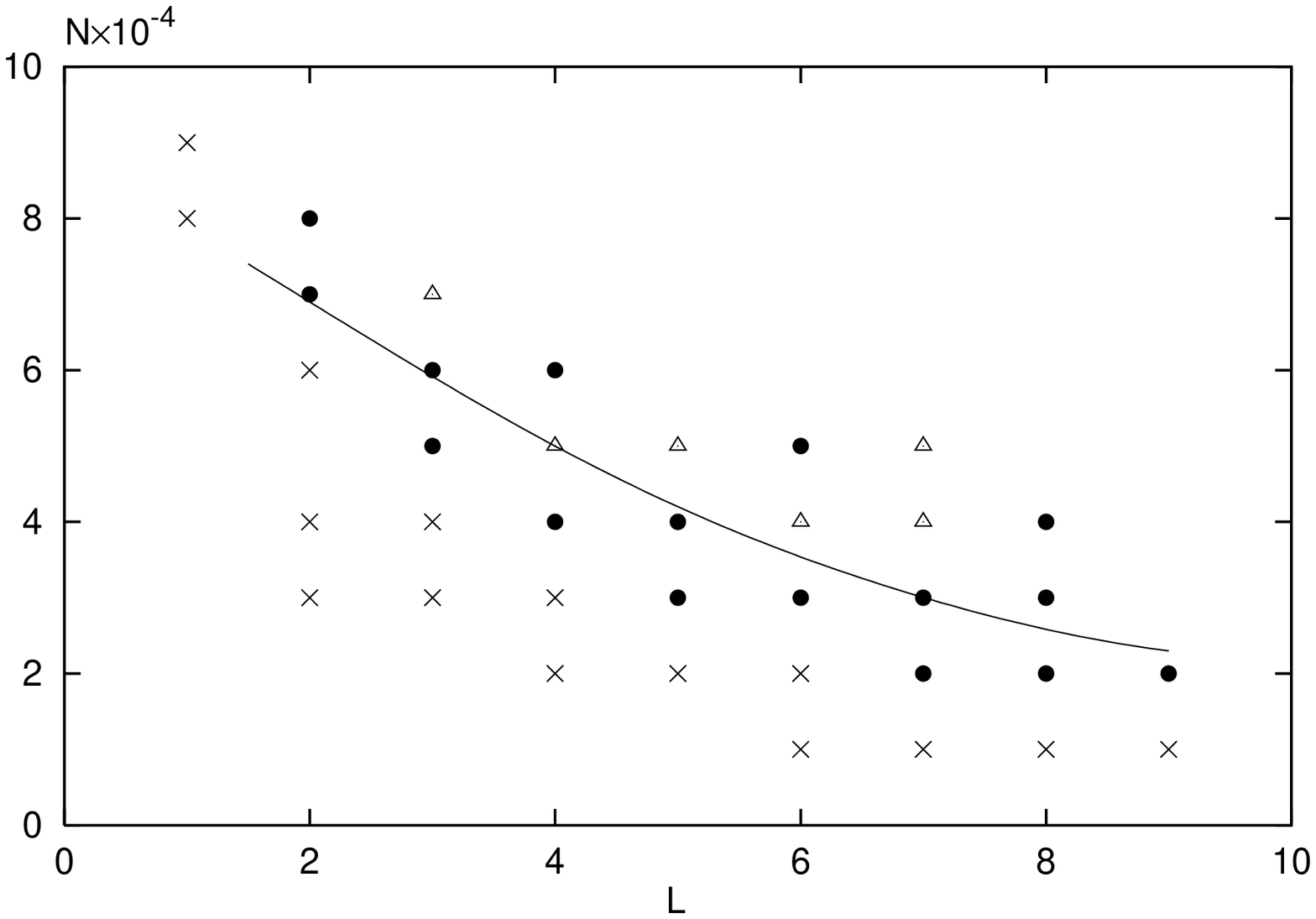}}} \par}
\vspace{0.3cm}
\caption{Phase diagram (L,N) for $M=3$. Different kinds of plots of $N_g(s)$ are marked 
with different signs. In the case of small $N$, a part of plot for large $s$ is a horizontal 
line at $N_g=1$, as in Fig.1; this kind of plot is marked by ($\times$). The scale-free
power functions, as in Figs. 2 and 3; this kind of plot is marked by ($\bullet$). Finally, for the largest
$N$, in many cases we observe a curve with a cutoff, as in Fig.4; this kind of plot is marked by ($\triangle$).
A hypothetical percolation threshold is marked by a solid
line.}
\end{figure}

\section{Discussion}
As it was noted in the Introduction, our results appear to be close to what could be expected 
for an uncorrelated percolation in Bethe lattice \cite{t1}. For $M=3$, the percolation threshold
is $p_c=1/(M-1)=0.5$. By symmetry, we can interchange occupied cells and empty cells. Then, the 
pore distribution for $p>p_c$ is equivalent to the cluster size distribution below the percolation
threshold, i.e. for $p<p_c$, where the giant component is absent. At the percolation threshold, 
we expect that the average number of strands at the network nodes is $p_cM=1.5$. Actually,
we obtain the average number of strands equal to 1.39, 1.48, 1.52 and 1.53 for $L=7$ and 
$N=1,2,3$ and $4\times 10^4$, respectively. This means that, keeping the analogy with the 
uncorrelated percolation in the Bethe lattice, we are close to the percolation threshold. 
Theoretical value of the exponent $\tau$ in the two-dimensional Bethe lattice is 
$2.5$\cite{t1}. Taking into account our numerical errors, this is not in 
contradiction with our evaluations. Moreover, the cluster size distribution above the percolation 
threshold in the Bethe lattice theory fulfils the equation $N_g(s,L)/N_g(s,L_c)\propto exp(-cs)$.
This formula is checked in Fig. 6 with data from Figs. 3 and 4. Again, the obtained data 
are not in contradiction with theory.

However, our network is neither uncorrelated, nor the Bethe lattice. Correlations appear
because the strands are allowed to move until finally attached to some other strands.
From this point of view, the system is analogous to the diffusion-limited aggregation (DLA),
where the density of the growing cluster decreases in time. The difference is that in DLA,
the cluster grows from some nucleus, which is absent in our simulation. On the other hand, 
loops are not prevented here, unlike in the Bethe lattice. The similarity of our results
to the latter model allows to suspect that in random structure near the percolation threshold 
the contribution to the cluster size distribution from accidental loops remains small.

The same calculations are performed also for the two-dimensional square lattice. Here the 
power-law like character of the pore size distribution is less convincing. It seems likely
that this additional difficulty is due to the fact that the triangular lattice is
closer to the off-lattice model than the square lattice. In general, the lattice introduces
numerical artifacts, which can be expected to disturb the system for small values of
$L$. In the limit of large $L$, it is only the anisotropy of the lattice which is different
from the ideal off-lattice limit. This numerically induced anisotropy allows for three
orientations for the triangular lattice and two orientations for the square lattice. In an 
off-lattice calculation, the number of orientations would be infinite. 

\begin{figure} 
\vspace{0.3cm} 
{\par\centering \resizebox*{9cm}{7cm}{\rotatebox{270}{\includegraphics{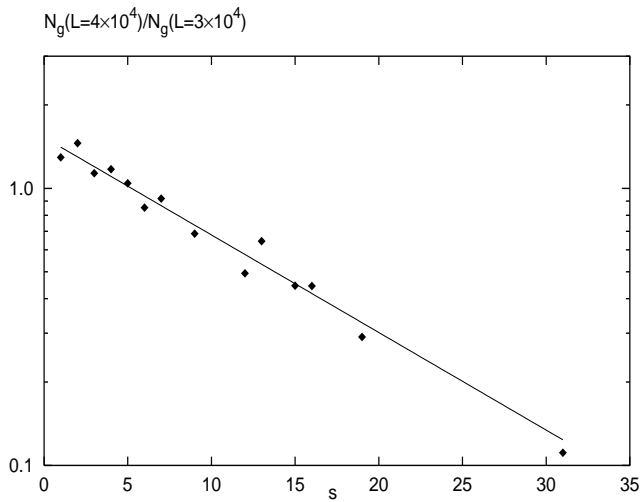}}} \par} 
\vspace{0.3cm} 
\caption{$N_g(L=4\times 10^4)/N_g(L=3\times 10^4)$ against s. The obtained points follow a 
straight line, in accordance with the theory of Bethe lattice \cite{t1}.}  
\end{figure}

\end{document}